\documentclass[letterpaper]{article}
\usepackage{aaai}
\usepackage{times}
\usepackage{helvet}
\usepackage{courier}
\usepackage{subfigure}
\usepackage{graphicx}
\usepackage{booktabs}
\begin{document}
\title{Searching for Representation: A sociotechnical audit of googling for members of U.S. Congress}
\author{Emma Lurie \\
 University of California, Berkeley \\
 \texttt{emma\_lurie@berkeley.edu} \\\And
 Deirdre K. Mulligan \\
 University of California, Berkeley \\
 \texttt{dmulligan@berkeley.edu} \\}
\maketitle

\begin{abstract}
\begin{quote} 
High-quality online civic infrastructure is increasingly critical for the success of democratic processes. There is a pervasive reliance on search engines to find facts and information necessary for political participation and oversight. We find that approximately 10\% of the top Google search results are likely to mislead California information seekers who use search to identify their congressional representatives. 70\% of the misleading results appear in featured snippets above the organic search results. We use both qualitative and quantitative methods to understand what aspects of the information ecosystem lead to this sociotechnical breakdown. Factors identified include Google's heavy reliance on Wikipedia, the lack of authoritative, machine parsable, high accuracy data about the identity of elected officials based on geographic location, and the search engine's treatment of under-specified queries. We recommend steps that Google can take to meet its stated commitment to providing high quality civic information, and steps that information providers can take to improve the legibility and quality of information about congressional representatives available to search algorithms. 
\end{quote}
\end{abstract}

\section{Introduction}

Search engines are an important part of the ``online civic infrastructure'' \cite{thorson2018political} that allows voters to access political information \cite{dutton2017social,sinclair2015googling}. While an informed citizenry is considered a prerequisite for an accountable democracy \cite{carpini1996americans}, political scientists have long understood that the cost of becoming a well-informed voter is too high for most Americans \cite{lupia2016uninformed}. Quality online civic infrastructure can reduce this cost, enabling individuals to locate information necessary for meaningful engagement in elections and other democratic processes. Many forms of civic participation 
require constituents to contact their member of Congress \cite{eckman2017constituent}. Contacts from non-constituents are often ignored, making it important for constituents to correctly identify their representatives. A 2017 report found that only 37\% of Americans know the name of their congressional representative.\footnote{https://www.haveninsights.com/just-37-percent-name-representative}
The 435 congressional districts are drawn using 9-digit zip codes, so districts often split counties, towns, and 5-digit zip codes.

Individuals turn to search engines to fill this information gap. The tweets below indicate such reliance and motivate our inquiry.

\begin{quote} 
\textit{She didn't know how to get in contact with a member of Congress? Um...Google?
}\end{quote}


\begin{quote}
\textit{@RepJayapal we urge all DEMs to call their Reps as we did and demand Nancy begin impeachment! It's very easy if you don't know who your rep is, google ``House of Representative'' then put in your zip code! Your rep \& their phone number will appear, CALL YOUR REP! @SpeakerPelosi \#ImpeachTrump}
\end{quote}

Approximately 90\% of U.S. search engine queries are performed on Google search.\footnote{https://gs.statcounter.com/search-engine-market-share/all/united-states-of-america} Thus Google's search results influence the relative visibility of elected officials and candidates for elected offices, and how they are presented and perceived \cite{diakopoulos2018vote}.

\begin{figure}
 \centering
 \includegraphics[width=0.4\textwidth]{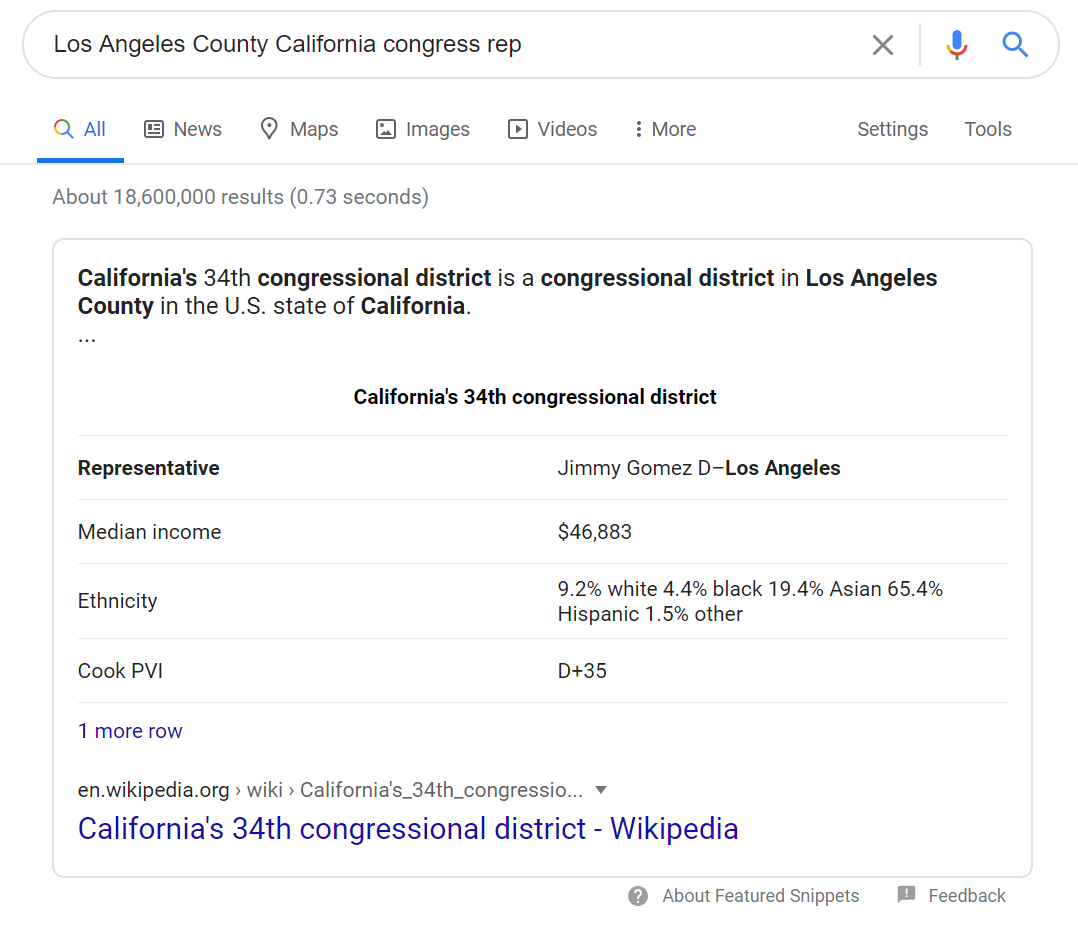}
 \caption{A screenshot of the featured snippet that appears when searching about LA County. Despite the county containing 18 distinct congressional districts, many SERPs surface this featured snippet that name Rep. Jimmy Gomez is the only representative of the county.}
 \label{fig:la_county}
\end{figure}

In response to a user query, Google returns a search engine results page (SERP). This paper is especially interested in the role of featured snippets in filling this information gap. Featured snippets are a Google feature that often appears as the top result and provides ``quick answers'' generated using ``content snippets'' from ``relevant websites'' \cite{googlesearchhelp} (see Figure~\ref{fig:la_county} for an example of a featured snippet). While the technical details of the feature are not public information, Google frames featured snippets as higher quality by placing them in the top position on the SERP and by instituting stricter content quality standards for featured snippets, and conveying those standards to the public. 


This study explores Google's search performance on the task of identifying the U.S. congressional representative associated with a geographic location. The search engine auditing community has evaluated how biases influence access to information about elections. Researchers have explored whether search engine biases can be exploited to manipulate elections \cite{epstein2015search} and the partisan biases of search results \cite{metaxas2017manipulation,Kliman-Silver:2015,Robertson:2018:APC:3178876.3186143}. We conduct an information seeker-centered search engine audit and find that approximately 10\% of SERPs surface a misleading top result in response to queries for California congressional representatives in a given location (e.g. ``San Diego rep name''). Both featured snippets and organic search results yield incorrect information; however, we focus on results that are both inaccurate and likely to mislead a reasonable information seeker to misidentify their congressperson.

The identity of the congressional representative for a specific street address or nine-digit zip code is unambiguous and can be retrieved from numerous web resources. Following elections, news sources and state election boards, report the name of all elected officials including members of Congress. Moreover, every member of Congress has an official house.gov site as well as a Wikipedia page. Given this, and Google's commitment to ``providing timely and authoritative information on Google Search to help voters understand, navigate, and participate in democratic processes,''\footnote{https://elections.google/civics-in-search/} search results for this basic, relatively static, factual information should be accurate. 

To answer the research question: \textit{what are the causes of the algorithmic breakdown?} we explore how various actors--Google, information providers, information seekers--as well as information quality contribute to this \textit{sociotechnical breakdown}. Looking at how the actors and their interactions align with Google's expectations provide insight into the sociotechnical nature of this breakdown as well as the various sites and options for intervention and repair. 


This paper offers the following contributions:

\begin{enumerate} 
\item We identify four distinct concepts as contributing to the breakdown: 1) Google's reliance on Wikipedia pages, 2) variations in relative algorithmic legibility and information completeness across information providers, 3) insufficiently specific queries by information seekers, and 4) interaction between the convoluted nature of congressional districts and challenges in geographic information retrieval. The sociotechnical analysis of this paper provides insight into the interactions between these factors.
\item We outline a novel method that combines a user-centered sock-puppet audit protocol \cite{mustafaraj2020case,hu-2019-www} with interpretivist methods to explore sociotechnical breakdown.
\item We provide recommendations for Google and other actors to improve the quality of search results for uncontested political information.

\end{enumerate}


\section{Related Research}

\subsection{Breakdown}
Winograd and Flores define the theoretical concept of breakdown as ``...a situation of non-obviousness, in which the recognition
that something is missing leads to unconcealing (generating through our declarations) some aspect of the network of tools that we are engaged in using'' \cite{winograd1986understanding}.

 As designers create new technical objects, they imagine how individuals will use the tools. \cite{winograd1986understanding,akrich1992scription}. 
Akrich argues that users' behavior frequently does not follow the script set up by designers, and breakdowns occur on unanticipated dimensions \cite{akrich1992scription}. Research has looked at breakdown as a means through which algorithms become visible to users in both the Facebook algorithm \cite{bucher2017algorithmic} and the Twitter algorithm \cite{burrell2019users}.

Previous research has examined the sociotechnical breakdown in Google search that occurs when content that denied the Holocaust surfaced in response to the query ``did the holocaust happen'' \cite{mulligan2018rescripting}. They propose the ``script'' \cite{akrich1992scription} of search ``reveals that perceived failure resides in a distinction and a gap between search results... and the results-of-search (the results of the entire query-to conception experience of conducting a search and interpreting search results)'' \cite{mulligan2018rescripting}. 


\subsection{Algorithm Auditing}
Due to the role search engines play in shaping individuals' lives, researchers are increasingly using audits to document their performance and politics. Sandvig et al. proposes algorithm audits as a research design to detect discrimination on online platforms \cite{sandvig2014auditing}. 
Previous audits of search engines in the political context have primarily concerned themselves with whether search engines are biased. While these studies have found limited evidence of search engine political bias \cite{Robertson:2018:APA:3290265.3274417}, there is some evidence of personalization of results by location \cite{Kliman-Silver:2015,Robertson:2018:APC:3178876.3186143} and a lack of information diversity in political contexts \cite{steiner2020seek}. 
Metaxa et al. recasts search results as ``search media'' and proposes longitudinal audits as a way to understand political trends \cite{metaxa2019search}. Other research has audited Google's non-organic search results finding that Google featured snippets amplify the partisanship of a SERP \cite{hu-2019-www}, Google Top Stories highlights primarily recent, mainstream news content \cite{trielli2019search}, Google autocomplete suggestions have a high churn rate \cite{robertson2019auditing}, and that Google knowledge panels for news sources are inconsistent and reliant on Wikipedia \cite{lurie2018investigating}.

Other studies have used a lens of algorithmic accountability to explore the impact of algorithms through algorithm audits \cite{diakopoulos2015algorithmic,raji2020closing,robertson2019auditing,steiner2020seek}. However, auditing is primarily used as a tool to provide transparency into the workings algorithms \cite{diakopoulos2015algorithmic}. The lens of breakdown , in contrast, centers the sociotechnical system exposing a broader set of actors who interact with and through the algorithm to scrutiny, and focuses the analysis on understanding what it means for an algorithmic system ``to work.''

\subsection{Politics \& Search Engines}
Search engines help facilitate voters' access to political information which supports engagement in democratic processes \cite{dutton2017social,sinclair2015googling,trevisan2018google}. Google has responded to information seekers' reliance on its platform for access to civic and political information by augmenting search results for political candidates, and other civic information \cite{diakopoulos2018vote}.

Researchers have raised concerns about biases and potential for manipulation \cite{epstein2015search}. Search engines, a type of online platform, are not neutral, apolitical intermediaries \cite{gillespie2010politics}. Previous research has sought to detect filter bubbles  by measuring search engine personalization \cite{pariser2011filter}. Information seekers specific query formulations \cite{kulshrestha2017quantifying,tripodi2018searching,Robertson:2018:APA:3290265.3274417} and the resulting suggestions \cite{bonart2019investigation,robertson2019auditing} shape search results; however others have not identified meaningful differences in searchers with different partisan affiliations query formulation \cite{trielli2020partisan}. Additionally, the politics of platforms are not limited to information seekers and platforms. Information providers are constantly working to become algorithmically recognizable to search engines \cite{gillespie2017algorithmically}, sometimes hijacking obscure search query formulations and filling data voids \cite{golebiewski2018data}. Google search results heavily favor Wikipedia \cite{vincent2021deeper,pradel2020biased} as well as Google operated results (e.g. non-organic search results) \cite{jeffries2020google}. 

\section{Algorithm Audit Study}

To understand the complexities of the online civic infrastructure, we perform a scraping audit, when an automated program makes repeated queries to a platform \cite{sandvig2014auditing}. To bolster our information seeker-centered perspective, our audit was informed by a survey to identify realistic query formulations. To support our interest in auditing the results-of-search, rather than the search results, we take an interpretivist approach to analyzing the audit results. This allows us to distinguish and focus on results that are likely to mislead information seekers, rather than the larger corpus of inaccurate search results.

\subsection{Modeling Information Seekers} 

\begin{table*}[]
\caption{Search terms selected to use for audit. We search these 26 queries terms for a sample of US zip codes, places (towns, cities, etc.), counties, and all CA congressional districts. These are the generalized queries extracted from user query formulations. The occurrence of each generalized query appears in parentheses.}
\label{tab:searchterms}
\begin{tabular}{p{3.5cm} p{13cm}}
\toprule
geographic granularity & query \\
\midrule
no specification & congressional representative (3); who is my congressional representative (2); my congressman (2); representative (2)\\

state & STATE congressional representatives (4); congressional representatives STATE (2);congressional representative STATE (2); STATE representatives (2); ]STATE congressional representative (2) \\

county & COUNTY house representative (2); COUNTY congressional representative (2); COUNTY STATE congress rep (1); my congressional representative COUNTY (1); who is my representative in COUNTY STATE (1) \\

place (e.g. city, town, CDP) & PLACE STATE congressional representative (2); PLACE congressional representative (2); congressional representative PLACE STATE (2); PLACE STATE congressman (2); PLACE ABBR congressman (1); PLACE ABBR representative (1); house of rep PLACE (1); congressional rep of PLACE STATE (1); PLACE congressperson (1) \\

zip code & congressional representatives STATE ZIP CODE(2); congress rep ZIP CODE (1); representative ZIP CODE (1); congressional representative ZIP CODE (1) \\
congressional district & STATE CD representative (1) \\
\end{tabular}
\end{table*}


It is important to use information seekers' query formulations to assess the performance of the search results. Typically researchers use 1) web browser search history data or 2) Google Trends for user query formulations. However, large-scale browser history data is typically not available for researchers who do not work at a search engine company. Installing a browser extension on users' computers is another technique employed to access users' query formulations, but this is an expensive method that invades user privacy, and for this study would involve massive over collection. Alternatively, Google Trends provides web users with a sample of real users' search queries. While high frequency queries may be well represented in the sample they are too low volume to register on Google Trends. That does not mean that no users formulate these queries, but the location-specific nature of these queries certainly make them less likely to register on Google Trends. Therefore, we follow Mustafaraj et al.'s \cite{mustafaraj2020case} approach of developing a survey to solicit user search terms. 

We obtained IRB approval and compensated 150 Mechanical Turk workers \$2 to answer the following survey question: 

\begin{quote}

``You are having an issue with your Social Security check. Your neighbor suggests that you contact your congressional representative's office to get the issue resolved. Use Google to search for your representative's name.''
\end{quote}

We chose this scenario as opposed to the advocacy examples in the motivating tweets because we wanted a scenario that is relatable to participants regardless of party affiliation or history of political engagement. Assistance navigating the federal bureaucracy is a standard constituent service offered by congressional offices, and one constituents consistently and broadly request. 

We asked participants to search for the name of the congressional representative, record the name of their representative, and then record all of the queries they searched to find the name of their representative. The 150 survey responses generated 166 related queries. Participants took on average 3.5 minutes to complete the survey. 16 queries appeared more than once when generalized (e.g. ``California rep name'' becomes ``STATE NAME rep name''). We used these 16 and selected 12 additional queries of various geographic granularity and shared similar query construction formats. See Table~\ref{tab:searchterms} for the full list of generic query terms.

The breadth of queries generated by users illustrates the lack of a common search strategy among participants. While some users searched by state name, others searched with some combination of county, place (e.g.town, city, community), (5-digit) zip code, congressional district, state name, and state abbreviation. 

We find that users search for the name of their representatives using various levels of geographic granularity that often do not provide a one-to-one mapping with congressional districts.

\subsection{Methods: Audit configuration}

This study focuses on data collected about California congressional representatives on two separate data collection rounds (May 3-5, 2020 and May 11-13, 2020). The two rounds allow us to 1) measure the variance in top results between the rounds and 2) verify that our results are not anomalous. We collected 1,803 SERPs in each data collection round that contain data about 146 zip codes, 117 places, and 36 counties in California.\footnote{Results from the vacant congressional districts (CA-25 and CA-50) are filtered out, as the information was in flux during the data collection period which coincided with either a special election or campaigning.} Each SERP is stored as an HTML file, but parsed with a modified version of the \texttt{WebSearcher}\footnote{https://github.com/gitronald/WebSearcher} library. The location parameter on the WebSearcher library makes all queries to Google appear to originate from the Bakersfield, California. 

We selected California for this paper because 1) it is the most populous state with 53 congressional districts; 2) there are both urban and rural communities and 3) California relies on an independent redistricting commission to draw congressional district boundaries. The commission has a mandate to keep communities, counties, cities and zip codes in the same congressional district where possible. So, we chose to analyze California data with the assumption that it would be one of the best performers for queries for congressional representatives (compared to states with histories of gerrymandered districts). Future work will extend this analysis to additional states.

\subsection{Audit Results} 

Table ~\ref{tab:auditresults} displays the results of data collection. 
The results reported are the top results. If the first result is a featured snippet, that is the top result. If the first result is an ``organic" search result, that is the top result. Over half of the top results are featured snippets, and of those 70\% are from Wikipedia. We focus the majority of our analysis on featured snippets, as they are, by Google's own standards, designed to provide a quick and more authoritative answer to information seekers.
Given the role Google intends featured snippets to play in guiding searchers to reliable information, the prevalence of misleading information about civic information is a breakdown of the online civic infrastructure worthy of interrogation.

Overall, 50\% of the top results are sourced from Wikipedia. Approximately 30\% of the top results are congressional representative's official house.gov websites (including officials sites and the house.gov ``Find your Representative'' tool. The remaining 20\% of top results are a combination of geographic look up tools (e.g. govtrack.us), county pages, and miscellaneous sites. 

Across the two rounds, there is high turnover in the content in the top result, yet consistent rates of misleading top results. The top result changes in 30\% of queries from Round 1 to Round 2. 12\% of featured snippets change between rounds. 11\% of results that were likely to mislead changed between rounds.


The Wikipedia articles surfaced in featured snippets vary. Some articles provide a detailed account of a geographic location, congressperson, or congressional district, others are incomplete, mentioning only one congressional representative in a location represented by multiple congressional districts, or omitting congressional representatives.

\subsection{Defining ``likely to mislead''}
\begin{figure}
\centering
 \hfill
 \subfigure[This top result appears when searching ``california representative.'' More than one representative is listed, so information seekers would likely have to do more research. ]{\includegraphics[width=0.4\textwidth]{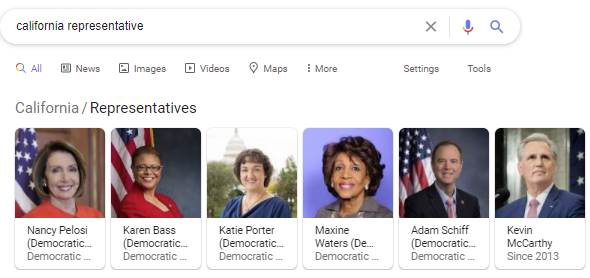}}
 \hfill
 \subfigure[The top result for the search result for ``representative 92317'' returns a job posting in the top result.]
 {\includegraphics[width=0.3\textwidth]{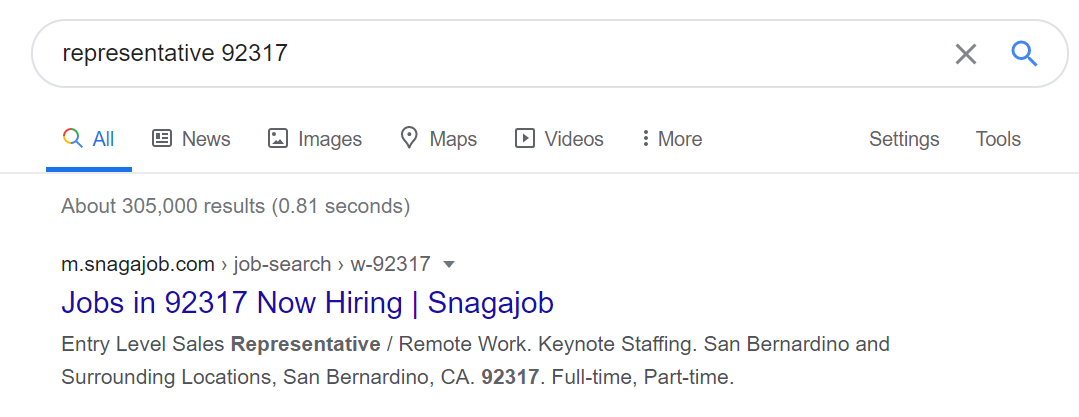}}
 \hfill
 \caption{These top results are incorrect, but do not mislead a reasonable information seeker to incorrectly identify their representative. We do not count these as misleading results.}
 \label{fig:incorrectnotmisleading}
\end{figure}

Centering the needs of information seekers requires us to consider not only how they construct queries, but how they interpret the search results revealed by the audit, construct the results of search \cite{mulligan2018rescripting}. While many of the results are technically inaccurate, they are not all equally likely to mislead information seekers. Counting any top result that does not contain the name of the correct congressional representative as incorrect would conflate inaccurate results with those likely to mislead an information seeker. To further illustrate, the top results in Figure~\ref{fig:incorrectnotmisleading} do not identify the correct congressional representative, yet we do not believe they are likely to mislead an information seeker as they either clearly signal a failed search or an ambiguous answer.


To support our interest in the results of search, we focus our analysis on top search results that are ``likely to mislead'': those that provide only a single congressional representative within the relevant state where that deterministic top result is either incorrect (i.e. the congressperson doesn't represent that region) or is incomplete (i.e. the region is represented by multiple representatives). This notion of whether top search results are ``likely to mislead'' is motivated from the U.S. Federal Trade Commission, where the standard for consumer deception includes testing for whether a reasonable consumer would be misled \cite{dingell1983ftc}.

As an example, if we are searching for L.A. County representative, and the top result says that Rep. Jimmy Gomez is the rep for L.A. County, this is considered a ``likely to mislead'' top result. While Rep. Gomez is one of 18 members of the House of Representatives to represent a portion of L.A. County, this answer is classified as likely to mislead because it likely leads the searcher to believe that their representative is Rep. Gomez.

\begin{table}[]

\caption{Approximately 12\% of Google SERPs return a misleading name of a congressperson in a featured snippet. Over half of these snippets are extracted from Wikipedia.}
\label{tab:auditresults}
\scalebox{0.65}{
  \begin{tabular}{llrr}
\toprule
   &                & Rd 1 (n=1803) & Rd 2 (n =1803)\\
\midrule
\textbf{featured snippet} & {} &   990 (55\%)&   899 (50\%) \\
   & misleading &   154 (16\%)&   149 (17\%) \\
   & Wikipedia &   680 (69\%)&   730 (81\%)\\
   & misleading \& Wikipedia &   107 (11\%) &   120 (13\%)\\
\textbf{other} &        &   813 (45\%) &   904 (50\%)\\
   & misleading &    57 (7\%) &    61 (7\%)\\
   & Wikipedia &   231 (28\%)&   223 (25\%)\\
   & misleading \& Wikipedia &    24 (3\%) &    16 (2\%) \\
\bottomrule
\textbf{likely to mislead}     & {}  & 211 (12\%) & 210 (12\%) \\
\textbf{Wikipedia}        & {}    & 911 (51\%) & 953 (53\%) \\
\textbf{misleading \& Wikipedia} & {} & 131 (7\%) & 136 (8\%) \\
\bottomrule
\end{tabular}}
\end{table}

After the authors agreed on the definition of ``likely to mislead,'' the first author did all of the labeling as there was limited opportunity for disagreement between labelers given the definition.

\subsection{Interrogating Algorithmic Breakdown} 

Previous research on the sociotechnical nature of search engines has generated several theories about how algorithmic ranking methodologies and platform practices, and the behavior of other actors--information providers, information seekers, advertisers--can contribute to sociotechnical breakdowns in algorithmically driven systems. We draw on these insights to identify potential sources that might contribute to the breakdown we identified.

One possible explanation is data voids. Golebiewski and boyd describe data voids as ``search terms for which the available relevant data is limited, non-existent, or deeply problematic.'' \cite{golebiewski2018data}. They identify five types of data voids: breaking news, strategic new terms, outdated terms, fragmented concepts, and problematic queries. While many of the queries we study can be considered obscure, the poor quality results we observe do not neatly fit into any of the five categories. Often with data voids, the search query is hijacked before authoritative content is created. Here, the authoritative content exists, yet it is not being surfaced on the SERP. Moreover, Golebiewski and boyd are concerned about intentional manipulators, but we find no evidence of intentional manipulation after examining the output for systematic output biases. 
In fact, we find no significant statistical difference in the  rates of misleading results across gender or party affiliation, or whether the zip code is in an urban vs. rural area. Many of the results arise from search results containing a partial list of representatives, and while the error rate differs by congressional representative, we manually reviewed the results of the representatives with the highest prevelance of misleading content and were unable to detect any evidence of efforts to depress results of a specific congressperson or political party (e.g. astroturfing, advertisements, different sources in the top position). 


While insights from previous audit research do not fully explain the poor performance, other known difficulties in information retrieval play a role. 

In particular, place name ambiguity plays an important role in the breakdown. Looking at the SERPs, one is struck by the number of misleading results concerning Los Angeles County. Los Angeles County is divided into 18 different congressional districts. However, when searching for LA County representatives, the top search result is frequently a featured snippet about Rep. Jimmy Gomez (see Figure~\ref{fig:la_county}). This featured snippet is likely to mislead reasonable searchers. The issue extends to other counties in California, including San Diego County and San Bernadino County. In each instance, the member of Congress who represents a section of the city that contains the county name (e.g. Los Angeles, San Diego, etc.) is disproportionately returned as the top result. This is consistent with prior research finding that information retrieval systems perform poorly when trying to discern location from ambiguous place names \cite{buscaldi2009toponym}.


\section{Analyzing the Platform} 


Google informs users that featured snippets are held to a different quality standard then organic search results. Google states that snippets are generated using ``content snippets'' from ``relevant websites.'' This indicates that with respect to snippets Google is exercising judgement about which sites are relevant to the ``quick answers" provided in a snippet, and either not or not exclusively relying on content identified through their standard metric of relevance \cite{googlesearchhelp}.

Google also informs users of the distinct content removal standards for snippets stating that ``featured snippets about public interest content -- including civic, medical, scientific and historical issues -- should not contradict well-established or expert consensus support'' \cite{googlesearchhelp}. Google writes that their ``automated systems are designed not to show featured snippets that don't follow our policies,'' but tells users that they ``rely on reports from our users'' to identify such content. Google indicates that they ``...manually remove any reported featured snippets if we find that they don't follow our policies'' and ``if our review shows that a website has other featured snippets that don't follow our policies or the site itself violates our webmaster guidelines, the site may no longer be eligible for featured snippets'' \cite{googlesearchhelp}. Users are asked to complain via the ``feedback'' button under the featured snippet (see Figure~\ref{fig:la_county}).

Google assigns itself the more modest task of drawing information from relevant websites. The meting out of responsibility here is noteworthy. Google only provides substantive quality standards for content removal, rather than inclusion. They have created a standard for other actors to police accuracy, but not a distinct quality benchmark they are trying to maintain with respect to ``consensus'' or accuracy. Google's script shifts responsibility to other stakeholders to maintain high-quality snippets on public interest content.

The uniqueness of the snippet quality standard is further underscored by Google's help page that includes content removal policies for featured snippets. Google explains that ``these policies only apply to what can appear as a featured snippet. They do not apply to web search listings nor cause those to be removed'' \cite{googlesearchhelp}.

The design and positioning of featured snippets also distinguishes them as more authoritative. Google states they ``receive unique formatting and positioning on Google Search and are often spoken aloud by the Google Assistant'' \cite{googlesearchhelp}. Google attributes the ``unique set of policies" to this special design and positioning. 

It is clear that Google does not want stakeholders to conflate featured snippets and organic search results, nor the distinct quality standards applied to each.

With respect to the queries at issue in our study, Google has made other relevant commitments. Google commits to ``providing timely and authoritative information on Google Search to help voters understand, navigate, and participate in democratic processes…''\footnote{\texttt{https://elections.google/civics-in-search/}} Researchers have discussed the various features that Google has augmented the SERP with in an attempt to more carefully curate information on elections, including issue guides, elected official knowledge panels, and more \cite{diakopoulos2018vote}. The identification of a congressional representative via search clearly falls under Google's commitment to help voters participate in democratic processes. Yet, currently, none of the additional scaffolding Diakopoulos et al. describes exists in our dataset. 


The results of the audit give us a sense of what sites Google views as ``relevant" for the snippets returned in response to user queries seeking congressional representatives. Wikipedia populates 50\% of top results in our data collection and 70\% of the featured snippets. Previous research \cite{mcmahon2017substantial,vincent2018examining,vincent2019measuring,vincent2021deeper} details the important role Wikipedia plays in improving the quality of Google search results, especially in domains that Google may otherwise struggle to surface relevant content. 
Setting aside the comparative accuracy of the information provided by these sites, discussed below, the question of which sites Google considers ``relevant'' for purposes of generating snippets is significant.


\section{Analyzing Information Providers} 
Three types of relevant information providers that display in the top position on the SERP are: 1) Wikipedia (\~50\% of top results), 2) representatives' official house.gov websites (\~30\% of top results), and 3) congressional representative geographic look-up tools (\~6\% of top results).

\subsection{Wikipedia}
Google's treatment of Wikipedia as ``relevant'' for snippets about congressional representatives contributes to this breakdown. Wikipedia does not exhaustively or equally cover all of the geographic areas or congressional districts. 

We do not identify any instances of Wikipedia articles listing the incorrect name of representatives within the Wikipedia article; however, some Wikipedia pages list only a subset of the representatives for the place. Article length and level of detail varies widely on the Wikipedia pages referenced. In particular, information about representatives for snippets is sourced from multiple categories of Wikipedia pages that have different purposes and inconsistent levels of detail. Some articles do not list a congressional representative. Other articles mention one of the congressional representatives in the district but not another. Congressional districts that split geographic locations appear to be a major source of problems. 

For example, the city of San Diego is part of five congressional districts. Yet, the Wikipedia page for the 52nd congressional district overwhelmingly appears as the featured snippet for several queries made about San Diego. As a result, Rep. Scott Peters of the 52nd congressional district, disproportionately appears in featured snippets about congressional representatives in San Diego, with no mention of the other representatives (similar to Figure~\ref{fig:la_county}) While Rep. Peters is \textit{one} of the congressional representatives for San Diego, he is not the only one. So, why does the Wikipedia page for the 52nd congressional district appear rather than the 53rd? We hypothesize that the answer lies in the first sentence of the article summary (which is often used as the majority of the text in featured snippets) for the 52nd congressional district:
\begin{quote} 
``The district is currently in \textbf{San Diego} County. It includes coastal and central portions of the city of \textbf{San Diego}, including neighborhoods such as Carmel Valley, La Jolla, Point Loma and Downtown \textbf{San Diego}; the \textbf{San Diego} suburbs of Poway and Coronado; and colleges such as University of California, \textbf{San Diego} (partial), Point Loma Nazarene, University of \textbf{San Diego}, and colleges of the \textbf{San Diego} Community College District.
\end{quote} 
``San Diego'' appears seven times in the above sentence. In comparison the Wikipedia page for the 53rd district, which encompasses sections of San Diego, only mentions San Diego twice in the first two sentences. While likely unintentional, the article summary for the Wikipedia page of the 52nd congressional district of California has effectively employed search engine optimization strategies to appear as the answer to queries about the congressional representative for San Diego. While the article summary explains the district is limited to coastal and central portions of San Diego, that distinction does not appear in the featured snippet. It seems unlikely that the contributors to that Wikipedia article were attempting to appear more `algorithmically recognizable' \cite{gillespie2017algorithmically} to search engines, but, that is the result.

To quantify possible coverage gaps in Wikipedia, we measured \textit{how many Wikipedia pages for California places (cities, towns, etc.) mention the name of at least one member of Congress?} To do this we randomly sampled 2,000 listed places from the child articles of the Wikipedia page ``List of Places in California''. We then searched for the presence of the name of a California member of Congress in the resulting articles. We found that less than half of Wikipedia pages (42\%) contained the name of a member of the California congressional delegation.

While increasing the level of detail on Wikipedia pages would likely improve overall search result quality, it is not only incorrect or even incomplete information on specific Wikipedia pages that contributes to the misleading content in snippets sourced from them. In short, Wikipedia articles are not constructed to provide accurate quick answers to these particular queries via featured snippets. While this is not inherently problematic on its own, given Google's well-documented reliance on Wikipedia, it is a contributing factor to the observed breakdown.


\subsection{Official Websites}

Every congressional representative has a house.gov hosted web page. These results appeared in the top result \~30\%, and comprise \~25\% of all misleading results. While many of the house.gov pages share a common template, the level of detail and modes of displaying district information varies by representative. Most websites have an ``our district'' page, that contains details about the congressional district. These pages almost always include an interactive map of the district. While this is a useful tool for information seekers who find their way to that page, the map is not `algorithmically recognizable' to Google. Some pages include a combination of counties, cities, zip codes, notable locations, economic hubs, images of landmarks. However, we found no correlation between the website design or content and better algorithmic performance for their district. 

\subsection{Geographic Lookup Tools} 

\begin{figure}
\centering
 \hfill

 \subfigure[House.gov Find My Rep Interface]{\includegraphics[width=.35\textwidth]{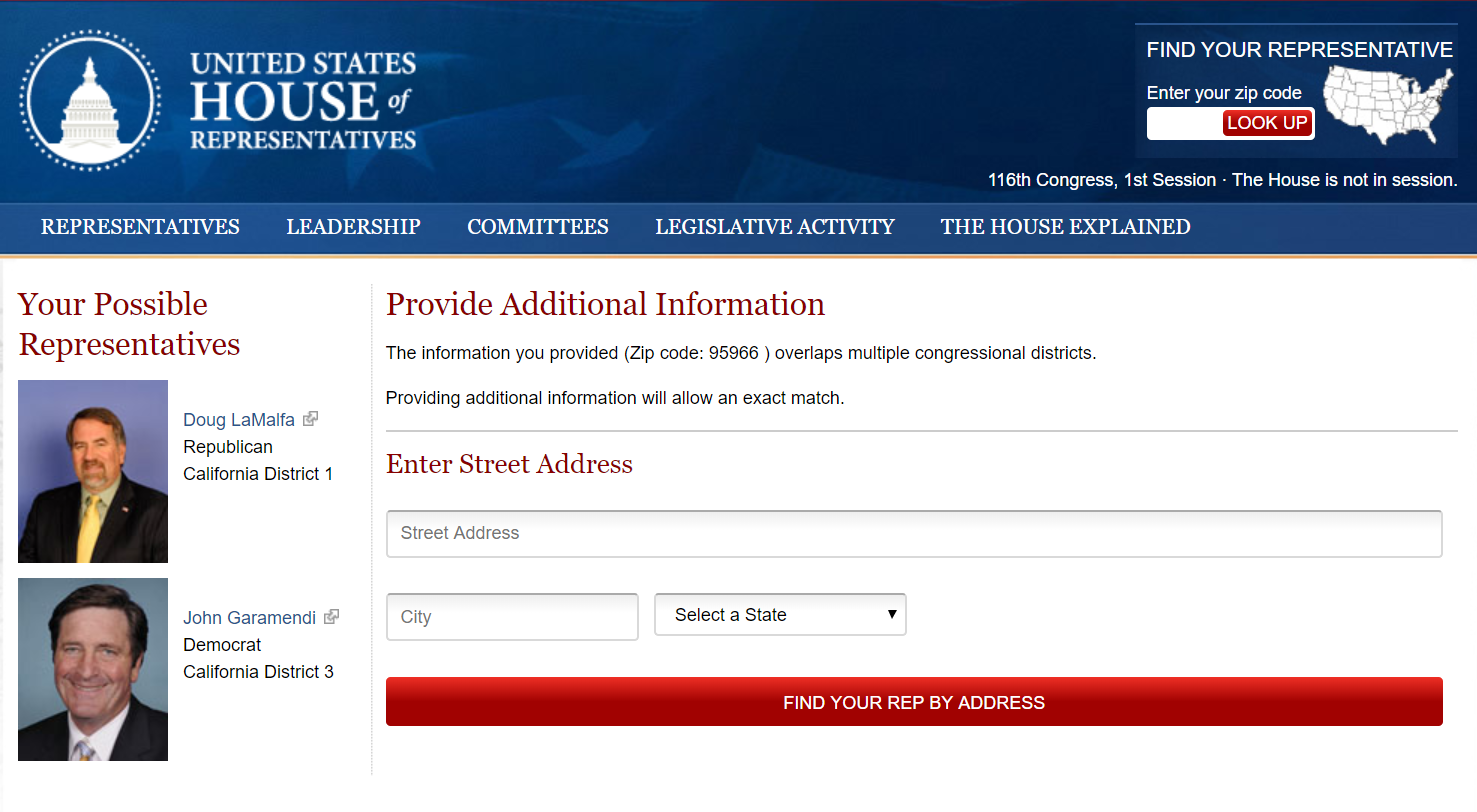}}
 \hfill
 \subfigure[Search results page for Find My Rep]{\includegraphics[width=.35\textwidth]{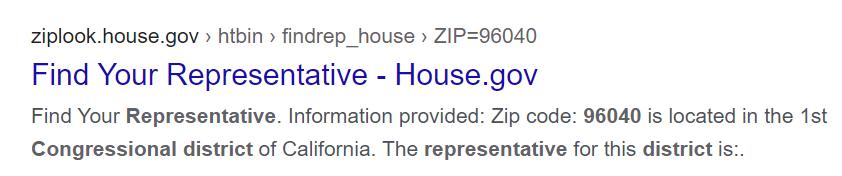}}
 \hfill
 \caption{Screenshots displaying house.gov's ``Find My Rep'' tool and the way it is displayed on the Google results page.}
 \label{fig:find_my_rep}
\end{figure}

A number of online tools help voters identify their representatives via geographic location. One such tool is house.gov's ``Find Your Representative'' tool (see Figure~\ref{fig:find_my_rep}). For 6\% of all SERPs, the house.gov ``find my rep'' tool appears in the top position in the organic search results. As shown in Figure~\ref{fig:find_my_rep}, the text snippet in the search result does not include the name of the representative. So, if information seekers want to identify the name of their representative, they must click on the search result. This limits the prevalence and usefulness of the authoritative information for searchers looking for a ``quick answer.'' 

However, the interface of the house.gov tool is different from other common representative look up tools, as its interface first asks information seekers to enter the zip code. If there is only one congressional representative in the five-digit zip code, the tool returns the name of the representative. If there are multiple congressional districts in the zip code, all the representatives in a zip code are displayed, and information seekers are then prompted to enter their street address. In contrast, the govtrack.us tool selects a set of geographic coordinates when you provide a zip code and Common Cause's ``Find your Representative tool'' requires a street address. 

\section{Analyzing Information Seekers} 
Google places responsibility for query formulation on information seekers. The long tail of the queries information seekers searched indicate that features like autocomplete are likely not directing information seekers to particular query constructions. Unlike related areas such as polling place look up where Google has provided forms to structure user queries to ensure correct results, Google has left information seekers to structure queries on their own. Our pre-audit information seeker survey finds that several common user query formulations do not specify geographic location or only specify the name of the state the information seeker is searching from. However, properly identifying an individual's congressional representative requires a nine-digit zip code or street address. Information seekers' queries are limiting the accuracy of the search results; however, in practice it may not be misleading searchers of the results of search as some inaccurate results likely trigger further information seeking rather than belief in an inaccurate answer. 

These under specified queries may indicate that information seekers are unfamiliar with how congressional districts are drawn. This assumption seems likely given the body of political science research that finds that Americans know little about their political system \cite{lupia2016uninformed,bartels1996uninformed,somin2006knowledge}. Additionally, a study of search queries information seekers use to decide who to vote for in the 2018 U.S. midterm elections (i.e. a non-presidential election) finds that a majority of voters search terms use ``low-information queries'' \cite{mustafaraj2020case} to find out more information about political candidates. 

Another explanation may be that users, like researchers, assume that Google uses location information latent in the web environment to refine search results. Given users' experiences with personalization, they may incorrectly assume that Google will leverage other sources of information about location to fill in the gap in their search query generated by their lack of political knowledge. 


\section{Limitations} 
This paper uses algorithmic measurement techniques (auditing) to understand the sociotechnical system of the online civic infrastructure in the United States. Future work may collect data for a broader geographic region or a collect longitudinal data, but these efforts are outside the scope and research question of this paper.

\textit{Survey design:} We asked participants in our survey to search until they identified their congressperson and then record their representative's name and all of the queries they searched. However when lists of representatives appeared as in Figure~\ref{fig:incorrectnotmisleading}, participants frequently selected the first name in the list (i.e. Nancy Pelosi). On one hand, this could indicate our criteria for content being ``likely to mislead'' information seekers is under-inclusive, as we assume that information seekers who are presented with multiple answers will seek out additional information rather than just selecting the first option on an image carousel (see Figure~\ref{fig:incorrectnotmisleading} for an example). A different interpretation is that participants weren't incentivized to identify the correct name of their representative (e.g. with a bonus). Recognizing this limitation, we do not conduct any analysis around iterative search behavior or analyze how many of our participants correctly identified the name of their representative. 
Further data on how users search may reveal additional methods and avenues for repair.


\textit{Data Collection:} This paper presents a subset of the data originally collected in the audit study. In an attempt to have the queries appear to be searched from different congressional districts (simulating users searching from home), we duplicated about 900 SERPs. Instead of the searches appearing to come from different congressional districts, they all originated in central California. As a result, we randomly select one copy of every unique query in our analysis (n=1803). There is additional work to do about the localization of the search results. Preliminary analyses find limited evidence of personalization, but more work is needed.

\section{Discussion: Contemplating Repair}

Our research identifies a breakdown created by relying on a generalized search script--relying on Wikipedia as a ``relevant site''--to shape search results in the domain of civic information. Introna and Nissenbaum critique the assumption that a single set of commercial norms should uniformly dictate the policies and practices of Web search \cite{introna2000shaping}. Nissenbaum also speculates that search engines will find little incentive to address more niche information needs, including those ``for information about the services of a local government authority'' \cite{nissenbaum2011contextual}. In particular, Introna and Nissenbaum emphasize how the power to drive traffic toward certain Web sites at the expense of others may lead those sites to atrophy or even disappear. Where the sites being made less visible are those that provide accurate information to support civic activity, often provided by public entities with public funds, the risks of atrophy or disappearance are troubling. Driving traffic toward Wikipedia rather than house.gov may depress support for public investment in online civic infrastructure. By elevating Wikipedia pages in response to search queries seeking the identity of congressional representatives Google may undercut investment in expertly produced, public election infrastructure in favor of a peer produced infrastructure that is not tailored to support civic engagement and is dependent on private donations.

\textit{Deemphaisizing Wikipedia:} One factor in the breakdown is Google's heavy reliance on Wikipedia to source featured snippets (70\% in our results). This pattern \cite{mcmahon2017substantial,vincent2018examining} is not unique to our queries, but it is problematic in the case of searches for elected officials. Based on our review of the of Wikipedia pages from which snippets are pulled, we believe Wikipedia is a poor choice of ``relevant site'' to meet the goal of aligning results on public interest topics with well-established or expert consensus in response to these queries. A sample of 1,196 working Wikipedia pages of places in California (e.g. city, town, unincorporated communities) finds that 58\% do not list the name of any congressional representatives. Others still only report one of the congressional representatives. The information on these pages is not constructed to help information seekers correctly identify their congressional representatives, especially not via extracted metadata or a partial article summary. 


\textit{Bespoke tools:} In other civic engagement areas Google has carved out an expanded role for itself in the search script. Rather than relying on seekers to craft effective queries, or relying on other sources to create accurate and machine legible content, Google has created structured forms, information sources, and APIs to fuel accurate results. Given relatively poor search literacy and political literacy, the current script assigns searchers too much responsibility for query formation. Given the public good of accurate civic information, and active efforts to mislead and misinform voters about basic aspects of election contests and procedures, such reliance on searchers poses risks. While we did not identify any activity to exploit the vulnerabilities we identify, this does not rule out such actions in the future. 

Google could create a new widget and display it on the top of the SERP for queries that signal a user's interest in identifying their congressional representative. Such a tool could augment or collaborate with existing resources like Google's Civic Information API or the house.gov ``Find your representative'' tool. There is no benefit to diverse results in this context. Allowing different sites to compete for users' eyeballs may encourage the proliferation of misinformation and indirectly to disenfranchisement. 

The house.gov interface balances information quality and privacy in an ideal way. Unlike other geographic look up tools like govtrack.us or the Google Civic Information API that default to request information seekers' street address, the house.gov tool first prompts information seekers for their zip code, and only prompts for their street address if their zip code overlaps with multiple congressional districts. The goal of a geographic look up tool that maps locations to congressional districts should be to accurately return data to all information seekers who navigate to the tool. Some information seekers may be skeptical of a tool that collects personal identifying information (e.g. street address), so providing information seekers with a less intrusive, but not always deterministic, initial prompt seems desirable. While geographic personalization is contextually appropriate to searches for congressional representatives, user studies to understand information seeker's experience and perception with geographic look-up tools in this context would aid design. 

Information providers can take independent steps to improve the `algorithmic recognizability' of information about congressional representatives for specific locations. Specifically, congressional officials house.gov pages can make their sites more legible to search engine crawlers. While over half of representative's home pages appear in at least one top result, they often appear in the top position for a small subset of queries about their district. The house.gov websites all contain an interactive map of the congressperson's district, which is useful for an information seeker already on that page, but is not meaningfully indexed by search engines. A substantial number of congressional representatives' websites are largely invisible to geographic queries. All House of Representatives web pages are managed by House Information Resources (HIR). Creating metadata that speaks directly to these queries would likely improve the performance of house.gov sites in snippets and organic search. The HIR could provide standards, or at the very least guidance, to members about web site construction aimed at improving the legibility of this information to search engines. While we do not propose one silver bullet solution, we view this as a solvable issue.

\section{Conclusion} 

 The online civic infrastructure is an emergent infrastructure composed of the sites and services that are assembled to support users’ civic information needs,  which is critical for meaningful participation in democratic processes.

There is a pervasive reliance on search engines to find facts and information necessary for civic participation. This research exposes a weakness in our current civic infrastructure and motivates future work in interrogating online search infrastructure beyond measuring partisan bias or information diversity of search results. Even under our stringent definition of misleading search results, we see real structural challenges in using search to provide information seekers with quick answers about uncontested political questions. Even in the high-stakes context of election information, this work finds deep similarities between work on the interdependence of Wikipedia and Google \cite{vincent2021deeper,mcmahon2017substantial}, particularly how Google sources many of their more detailed search results through Wikipedia. While we do not attempt to assess the magnitude of the harm of returning misleading information about representative identity to constituents, this is an open question that involves sociopolitical considerations in addition to the technical.

Google has recognized their role in this ecosystem, endeavoring to ensure that information seekers find accurate information about these issues, and has built out infrastructure on the SERP as well as public-facing APIs toward this end. Constituent searches for the identify of their member of Congress deserve more careful consideration. 



\bibliographystyle{aaai}
\bibliography{sample}

\begin{thebibliography}{}

\bibitem[\protect\citeauthoryear{Akrich}{1992}]{akrich1992scription}
Akrich, M.
\newblock 1992.
\newblock The de-scription of technical objects.

\bibitem[\protect\citeauthoryear{Bartels}{1996}]{bartels1996uninformed}
Bartels, L.~M.
\newblock 1996.
\newblock Uninformed votes: Information effects in presidential elections.
\newblock {\em AJPS}  194--230.

\bibitem[\protect\citeauthoryear{Bonart \bgroup et al\mbox.\egroup
  }{2019}]{bonart2019investigation}
Bonart, M.; Samokhina, A.; Heisenberg, G.; and Schaer, P.
\newblock 2019.
\newblock An investigation of biases in web search engine query suggestions.
\newblock {\em Online Information Review}.

\bibitem[\protect\citeauthoryear{Bucher}{2017}]{bucher2017algorithmic}
Bucher, T.
\newblock 2017.
\newblock The algorithmic imaginary: exploring the ordinary affects of facebook
  algorithms.
\newblock {\em Information, communication \& society} 20(1):30--44.

\bibitem[\protect\citeauthoryear{Burrell \bgroup et al\mbox.\egroup
  }{2019}]{burrell2019users}
Burrell, J.; Kahn, Z.; Jonas, A.; and Griffin, D.
\newblock 2019.
\newblock When users control the algorithms: Values expressed in practices on
  twitter.
\newblock {\em CSCW} 3(CSCW):1--20.

\bibitem[\protect\citeauthoryear{Buscaldi}{2009}]{buscaldi2009toponym}
Buscaldi, D.
\newblock 2009.
\newblock Toponym ambiguity in geographical information retrieval.
\newblock In {\em SIGIR},  847--847.

\bibitem[\protect\citeauthoryear{Carpini and
  Keeter}{1996}]{carpini1996americans}
Carpini, M. X.~D., and Keeter, S.
\newblock 1996.
\newblock {\em What Americans know about politics and why it matters}.
\newblock Yale University Press.

\bibitem[\protect\citeauthoryear{Diakopoulos \bgroup et al\mbox.\egroup
  }{2018}]{diakopoulos2018vote}
Diakopoulos, N.; Trielli, D.; Stark, J.; and Mussenden, S.
\newblock 2018.
\newblock I vote for—how search informs our choice of candidate.
\newblock {\em Digital Dominance: The Power of Google, Amazon, Facebook, and
  Apple, M. Moore and D. Tambini (Eds.)} 22.

\bibitem[\protect\citeauthoryear{Diakopoulos}{2015}]{diakopoulos2015algorithmic}
Diakopoulos, N.
\newblock 2015.
\newblock Algorithmic accountability: Journalistic investigation of
  computational power structures.
\newblock {\em Digital journalism} 3(3):398--415.

\bibitem[\protect\citeauthoryear{Dingell}{1983}]{dingell1983ftc}
Dingell, J.~D.
\newblock 1983.
\newblock Ftc policy statement on deception.

\bibitem[\protect\citeauthoryear{Dutton \bgroup et al\mbox.\egroup
  }{2017}]{dutton2017social}
Dutton, W.~H.; Reisdorf, B.; Dubois, E.; and Blank, G.
\newblock 2017.
\newblock Social shaping of the politics of internet search and networking:
  Moving beyond filter bubbles, echo chambers, and fake news.

\bibitem[\protect\citeauthoryear{Eckman}{2017}]{eckman2017constituent}
Eckman, S.~J.
\newblock 2017.
\newblock Constituent services: Overview and resources. congressional research
  service.

\bibitem[\protect\citeauthoryear{Epstein and
  Robertson}{2015}]{epstein2015search}
Epstein, R., and Robertson, R.~E.
\newblock 2015.
\newblock The search engine manipulation effect (seme) and its possible impact
  on the outcomes of elections.
\newblock {\em Proceedings of the National Academy of Sciences}
  112(33):E4512--E4521.

\bibitem[\protect\citeauthoryear{Gillespie}{2010}]{gillespie2010politics}
Gillespie, T.
\newblock 2010.
\newblock The politics of ‘platforms’.
\newblock {\em New media \& society} 12(3):347--364.

\bibitem[\protect\citeauthoryear{Gillespie}{2017}]{gillespie2017algorithmically}
Gillespie, T.
\newblock 2017.
\newblock Algorithmically recognizable: Santorum’s google problem, and
  google’s santorum problem.
\newblock {\em Information, communication \& society} 20(1):63--80.

\bibitem[\protect\citeauthoryear{Golebiewski and {boyd,
  d}}{2018}]{golebiewski2018data}
Golebiewski, M., and {boyd, d}.
\newblock 2018.
\newblock Data voids: Where missing data can easily be exploited.
\newblock {\em Data \& Society} 29.

\bibitem[\protect\citeauthoryear{{Google Search Help}}{2019}]{googlesearchhelp}
{Google Search Help}.
\newblock 2019.
\newblock How google's featured snippets work.
\newblock https://support.google.com/websearch/answer/9351707?hl=en.

\bibitem[\protect\citeauthoryear{Hu \bgroup et al\mbox.\egroup
  }{2019}]{hu-2019-www}
Hu, D.; Jiang, S.; Robertson, R.~E.; and Wilson, C.
\newblock 2019.
\newblock {Auditing the Partisanship of Google Search Snippets}.
\newblock In {\em {WWW 2019}}.

\bibitem[\protect\citeauthoryear{Introna and
  Nissenbaum}{2000}]{introna2000shaping}
Introna, L.~D., and Nissenbaum, H.
\newblock 2000.
\newblock Shaping the web: Why the politics of search engines matters.
\newblock {\em The information society} 16(3):169--185.

\bibitem[\protect\citeauthoryear{Jeffries and Yin}{}]{jeffries2020google}
Jeffries, A., and Yin, L.
\newblock Google’s top search result? surprise! it's google.
\newblock {\em The MarkUp}.

\bibitem[\protect\citeauthoryear{Kliman-Silver \bgroup et al\mbox.\egroup
  }{2015}]{Kliman-Silver:2015}
Kliman-Silver, C.; Hannak, A.; Lazer, D.; Wilson, C.; and Mislove, A.
\newblock 2015.
\newblock Location, location, location: The impact of geolocation on web search
  personalization.
\newblock In {\em IMC '15}, IMC,  121--127.
\newblock New York, NY, USA: ACM.

\bibitem[\protect\citeauthoryear{Kulshrestha \bgroup et al\mbox.\egroup
  }{2017}]{kulshrestha2017quantifying}
Kulshrestha, J.; Eslami, M.; Messias, J.; Zafar, M.~B.; Ghosh, S.; Gummadi,
  K.~P.; and Karahalios, K.
\newblock 2017.
\newblock Quantifying search bias: Investigating sources of bias for political
  searches in social media.
\newblock In {\em CSCW' 17},  417--432.
\newblock ACM.

\bibitem[\protect\citeauthoryear{Lupia}{2016}]{lupia2016uninformed}
Lupia, A.
\newblock 2016.
\newblock {\em Uninformed: Why people know so little about politics and what we
  can do about it}.
\newblock Oxford University Press.

\bibitem[\protect\citeauthoryear{Lurie and
  Mustafaraj}{2018}]{lurie2018investigating}
Lurie, E., and Mustafaraj, E.
\newblock 2018.
\newblock Investigating the effects of google's search engine result page in
  evaluating the credibility of online news sources.
\newblock In {\em Web Sci'18},  107--116.

\bibitem[\protect\citeauthoryear{McMahon, Johnson, and
  Hecht}{2017}]{mcmahon2017substantial}
McMahon, C.; Johnson, I.; and Hecht, B.
\newblock 2017.
\newblock The substantial interdependence of wikipedia and google: A case study
  on the relationship between peer production communities and information
  technologies.
\newblock In {\em AAAI ICWSM}.

\bibitem[\protect\citeauthoryear{Metaxa \bgroup et al\mbox.\egroup
  }{2019}]{metaxa2019search}
Metaxa, D.; Park, J.~S.; Landay, J.~A.; and Hancock, J.
\newblock 2019.
\newblock Search media and elections: A longitudinal investigation of political
  search results.
\newblock {\em CSCW 2019} 3(CSCW):1--17.

\bibitem[\protect\citeauthoryear{Metaxas and
  Pruksachatkun}{2017}]{metaxas2017manipulation}
Metaxas, P.~T., and Pruksachatkun, Y.
\newblock 2017.
\newblock Manipulation of search engine results during the 2016 us
  congressional elections.
\newblock In {\em ICIW 2017}.

\bibitem[\protect\citeauthoryear{Mulligan and
  Griffin}{2018}]{mulligan2018rescripting}
Mulligan, D.~K., and Griffin, D.~S.
\newblock 2018.
\newblock Rescripting search to respect the right to truth.

\bibitem[\protect\citeauthoryear{Mustafaraj, Lurie, and
  Devine}{2020}]{mustafaraj2020case}
Mustafaraj, E.; Lurie, E.; and Devine, C.
\newblock 2020.
\newblock The case for voter-centered audits of search engines during political
  elections.
\newblock In {\em Proceedings of FAccT 2020},  559--569.

\bibitem[\protect\citeauthoryear{Nissenbaum}{2011}]{nissenbaum2011contextual}
Nissenbaum, H.
\newblock 2011.
\newblock A contextual approach to privacy online.
\newblock {\em Daedalus} 140(4):32--48.

\bibitem[\protect\citeauthoryear{Pariser}{2011}]{pariser2011filter}
Pariser, E.
\newblock 2011.
\newblock {\em The filter bubble: What the Internet is hiding from you}.
\newblock Penguin UK.

\bibitem[\protect\citeauthoryear{Pradel}{2020}]{pradel2020biased}
Pradel, F.
\newblock 2020.
\newblock Biased representation of politicians in google and wikipedia search?
  the joint effect of party identity, gender identity and elections.
\newblock {\em Political Communication}  1--32.

\bibitem[\protect\citeauthoryear{Raji and Smart}{2020}]{raji2020closing}
Raji, I.~D., and Smart, A.
\newblock 2020.
\newblock Closing the ai accountability gap: defining an end-to-end framework
  for internal algorithmic auditing.
\newblock In {\em Proceedings of the FAccT 2020},  33--44.

\bibitem[\protect\citeauthoryear{Robertson \bgroup et al\mbox.\egroup
  }{2018}]{Robertson:2018:APA:3290265.3274417}
Robertson, R.~E.; Jiang, S.; Joseph, K.; Friedland, L.; Lazer, D.; and Wilson,
  C.
\newblock 2018.
\newblock Auditing partisan audience bias within google search.
\newblock {\em Proc. ACM Hum.-Comput. Interact.} 2(CSCW):1--22.

\bibitem[\protect\citeauthoryear{Robertson \bgroup et al\mbox.\egroup
  }{2019}]{robertson2019auditing}
Robertson, R.~E.; Jiang, S.; Lazer, D.; and Wilson, C.
\newblock 2019.
\newblock Auditing autocomplete: Suggestion networks and recursive algorithm
  interrogation.
\newblock In {\em ACM WebSci},  235--244.

\bibitem[\protect\citeauthoryear{Robertson, Lazer, and
  Wilson}{2018}]{Robertson:2018:APC:3178876.3186143}
Robertson, R.~E.; Lazer, D.; and Wilson, C.
\newblock 2018.
\newblock Auditing the personalization and composition of politically-related
  search engine results pages.
\newblock In {\em Proceedings of the 2018 World Wide Web Conference}, WWW '18,
  955--965.
\newblock Republic and Canton of Geneva, Switzerland: International World Wide
  Web Conferences Steering Committee.

\bibitem[\protect\citeauthoryear{Sandvig \bgroup et al\mbox.\egroup
  }{2014}]{sandvig2014auditing}
Sandvig, C.; Hamilton, K.; Karahalios, K.; and Langbort, C.
\newblock 2014.
\newblock Auditing algorithms: Research methods for detecting discrimination on
  internet platforms.
\newblock {\em Data and discrimination: converting critical concerns into
  productive inquiry} 22.

\bibitem[\protect\citeauthoryear{Sinclair and
  Wray}{2015}]{sinclair2015googling}
Sinclair, B., and Wray, M.
\newblock 2015.
\newblock Googling the top two: Information search in california’s top two
  primary.
\newblock {\em California Journal of Politics and Policy} 7(1).

\bibitem[\protect\citeauthoryear{Somin}{2006}]{somin2006knowledge}
Somin, I.
\newblock 2006.
\newblock Knowledge about ignorance: New directions in the study of political
  information.
\newblock {\em Critical Review} 18(1-3):255--278.

\bibitem[\protect\citeauthoryear{Steiner \bgroup et al\mbox.\egroup
  }{2020}]{steiner2020seek}
Steiner, M.; Magin, M.; Stark, B.; and Gei{\ss}, S.
\newblock 2020.
\newblock Seek and you shall find? a content analysis on the diversity of five
  search engines’ results on political queries.
\newblock {\em Information, Communication \& Society}  1--25.

\bibitem[\protect\citeauthoryear{Thorson, Xu, and
  Edgerly}{2018}]{thorson2018political}
Thorson, K.; Xu, Y.; and Edgerly, S.
\newblock 2018.
\newblock Political inequalities start at home: Parents, children, and the
  socialization of civic infrastructure online.
\newblock {\em Political Communication} 35(2):178--195.

\bibitem[\protect\citeauthoryear{Trevisan \bgroup et al\mbox.\egroup
  }{2018}]{trevisan2018google}
Trevisan, F.; Hoskins, A.; Oates, S.; and Mahlouly, D.
\newblock 2018.
\newblock The google voter: search engines and elections in the new media
  ecology.
\newblock {\em Information, Communication \& Society} 21(1):111--128.

\bibitem[\protect\citeauthoryear{Trielli and
  Diakopoulos}{2019}]{trielli2019search}
Trielli, D., and Diakopoulos, N.
\newblock 2019.
\newblock Search as news curator: The role of google in shaping attention to
  news information.
\newblock In {\em CHI},  1--15.

\bibitem[\protect\citeauthoryear{Trielli and
  Diakopoulos}{2020}]{trielli2020partisan}
Trielli, D., and Diakopoulos, N.
\newblock 2020.
\newblock Partisan search behavior and google results in the 2018 us midterm
  elections.
\newblock {\em Information, Communication \& Society}  1--17.

\bibitem[\protect\citeauthoryear{Tripodi}{2018}]{tripodi2018searching}
Tripodi, F.
\newblock 2018.
\newblock Searching for alternative facts.
\newblock {\em Media Manipulation Research Initiative. New York, NY: Data \&
  Society Research Institute.}

\bibitem[\protect\citeauthoryear{Vincent and Hecht}{2021}]{vincent2021deeper}
Vincent, N., and Hecht, B.
\newblock 2021.
\newblock A deeper investigation of the importance of wikipedia links to search
  engine results.
\newblock {\em Proceedings of the ACM on Human-Computer Interaction}
  5(CSCW1):1--15.

\bibitem[\protect\citeauthoryear{Vincent \bgroup et al\mbox.\egroup
  }{2019}]{vincent2019measuring}
Vincent, N.; Johnson, I.; Sheehan, P.; and Hecht, B.
\newblock 2019.
\newblock Measuring the importance of user-generated content to search engines.
\newblock In {\em Proceedings of the International AAAI Conference on Web and
  Social Media}, volume~13,  505--516.

\bibitem[\protect\citeauthoryear{Vincent, Johnson, and
  Hecht}{2018}]{vincent2018examining}
Vincent, N.; Johnson, I.; and Hecht, B.
\newblock 2018.
\newblock Examining wikipedia with a broader lens: Quantifying the value of
  wikipedia's relationships with other large-scale online communities.
\newblock In {\em CHI 2018},  1--13.

\bibitem[\protect\citeauthoryear{Winograd, Flores, and
  Flores}{1986}]{winograd1986understanding}
Winograd, T.; Flores, F.; and Flores, F.~F.
\newblock 1986.
\newblock {\em Understanding computers and cognition: A new foundation for
  design}.
\newblock Intellect Books.

\end{thebibliography}

\end{document}